\documentclass[12pt]{iopart}
\usepackage{epsfig}

\newcommand{\be}{\begin{equation}}
\newcommand{\ee}{\end{equation}}
\newcommand{\bea}{\begin{eqnarray}}
\newcommand{\eea}{\end{eqnarray}}
\newcommand{\beastar}{\begin{eqnarray*}}
\newcommand{\eeastar}{\end{eqnarray*}}
\newcommand{\nn}{\nonumber\\}
\newcommand{\lav}{\left\langle}
\newcommand{\rav}{\right\rangle}

\newcommand{\eq}[1]{~(\ref{#1})}
\newcommand{\eqq}[2]{~(\ref{#1},\ref{#2})}
\newcommand{\eqqq}[3]{~(\ref{#1},\ref{#2},\ref{#3})}

\newcommand{\order}{{{\mathcal O}}}

\newcommand{\ie}{{\it i.e.}}
\newcommand{\eg}{{\it e.g.}}
\newcommand{\ar}{\leftarrow}

\newcommand{\tw}{t_{\rm w}}

\newcommand{\var}{\Delta^2(E)}
\newcommand{\dt}{\Delta t\,}
\newcommand{\dd}[1]{\frac{\partial}{\partial#1}}
\newcommand{\Tg}{T_{\rm g}}
\newcommand{\BM}{Barrat and M\'ezard}
\newcommand{\zt}{\zeta}

\begin{document}

\title{Fluctuation-dissipation relations in trap models}

\author{Peter Sollich\footnote[1]{Email peter.sollich@kcl.ac.uk}}

\address{Department of Mathematics, King's College London, London WC2R
2LS, UK}

\begin{abstract}
Trap models are intuitively appealing and often solvable models of
glassy dynamics. In particular, they have been used to study aging and
the resulting out-of-equilibrium fluctuation-dissipation relations
between correlations and response functions. In this note I show
briefly that one such relation, first given by Bouchaud and Dean, is
valid for a general class of mean-field trap models: it relies only on
the way a perturbation affects the transition rates, but is
independent of the distribution of trap depths and the form of the
unperturbed transition rates, and holds for all observables that are
uncorrelated with the energy. The model with Glauber dynamics and an
exponential distribution of trap depths, as considered by \BM, does
not fall into this class if the perturbation is introduced in the
natural way by shifting all trap energies. I show that
a similar relation between response and correlation nevertheless
holds for the out-of-equilibrium dynamics at low temperatures. The results
points to intriguing parallels between trap models with energetic and
entropic barriers.
\end{abstract}





\section{Introduction}

Trap models consist of a single particle, or equivalently an ensemble
of non-interacting particles, hopping in a landscape of traps of
energy $E$. Such models have been studied extensively and shown to
account qualitatively for many interesting features of glassy
dynamics, see
\eg~\cite{Bouchaud92,BarMez95,Dyre87,BouDea95,BouComMon95,MonBou96,%
Head00,BerBou02,FieSol02,Ritort03}. In the simplest case the rates for
transitions from one trap to another depend only on the energies of
the two traps. One then has a mean-field trap model, where no
information on any spatial organization is retained. This is the case
that will concern us here; for work on spatial trap models see
\eg~\cite{HauKeh87,RinMaaBou00,RinMaaBou01,BerBou03}.

In this paper I focus on the behaviour of trap models after a quench
into the glassy phase, and in particular on two-time correlation and
response functions. For a generic observable $m$ the correlation
function is $C(t,\tw)=\lav m(t)m(\tw)\rav - \lav m(t)\rav \lav
m(\tw)\rav$, while the (linear) response function $\chi(t,\tw)$
measures the change in $\lav m(t)\rav$ due to a conjugate field $h$
that is switched on at time $\tw<t$. The time of preparation of the
system by quenching is taken as the zero of the time axis, so that the
waiting time $\tw$ can alternatively be thought of as the ``age''
of the system at the time when the field is applied. Over the last
decade it has been recognized that out-of-equilibrium
fluctuation-dissipation (FD) relations between such correlation and
response functions are extremely useful for characterizing glassy
dynamics~\cite{CugKur93,CugKur94,CugKurPel97,CriRit02b}. One defines a
fluctuation-dissipation theorem (FDT) ``violation factor'' $X(t,\tw)$
by
\be
-\frac{\partial}{\partial \tw}\chi(t,\tw) = \frac{X(t,\tw)}{T}
\frac{\partial}{\partial \tw} C(t,\tw)
\label{Xdef}
\ee
so that $X=1$ corresponds to the usual equilibrium FDT. The value of
$X$ can be read off from the slope of a parametric ``FD plot'' of
$\chi$ versus $C$; see~\cite{FieSol02,SolFieMay02} for a discussion of
the effects of using either $t$ or $\tw$ as curve parameter. The
quantity on the l.h.s.\ of\eq{Xdef}, denoted $R(t,\tw)$ below, is the
response to a short field impulse at time $\tw$.

In glassy systems one typically finds that the decay of two-time
correlation functions $C(t,\tw)$ exhibits several regimes: an initial
decay to a plateau, with further relaxation taking place only on
``aging'' timescales $t-\tw$ that grow with the age $\tw$, for example
$t-\tw =\order(\tw)$. In mean-field spin
glasses~\cite{CugKur93,CugKur94,CugKurPel97} one finds that, in the
limit of large $\tw$, $X$ has well-defined and distinct values in
these regimes, corresponding to an FD plot made up of two straight
line segments: $X=1$ in the short-time regime, corresponding to
quasi-equilibrium, and $X<1$ in the aging regime.  In the latter, one
can then define an {\em effective temperature} by $T_{\rm eff}=T/X$.
This has been shown to have many of the properties of a thermodynamic
temperature~\cite{CugKur93,CugKur94,CugKurPel97}, opening up the
exciting prospect of an effective equilibrium description of
out-of-equilibrium dynamics.

The existence of effective temperatures in systems other than the now
canonical (\eg\ spherical $p$-) spin glass models has been the subject
of much research in recent years, but a coherent picture has yet to
emerge~\cite{CriRit02b}. In Bouchaud's trap model~\cite{Bouchaud92},
intriguing results have recently been found~\cite{FieSol02}: even
though the correlation functions $C(t,\tw)$ decay within a single
aging ``time sector'' $t-\tw=\order(\tw)$, the FDT violation factor
$X$ is not constant as one might expect by analogy with mean-field
spin glass models.  Instead it varies continuously with $(t-\tw)/\tw$,
resulting in a curved FD plot with an asymptotic slope (for
$(t-\tw)/\tw\to\infty$, \ie\ $C\to 0$) of $X_\infty=0$.

In the Bouchaud trap model glassy dynamics arises from the presence of
{\em energy} barriers. The aim of this paper is to analyse the FD
relations in a different trap model, due to \BM~\cite{BarMez95}, where
glassiness instead results from the presence of {\em entropic}
barriers. Because of the different physical mechanisms causing the
out-of-equilibrium behaviour, it is then not {\em a priori} clear
whether the FD relations of the two models should be related. I find
that some important aspects of the FD relations are indeed the same,
pointing to intriguing parallels between models with energetic and
entropic barriers that deserve to be explored further.

Very recently, Ritort~\cite{Ritort03} has also considered FD relations
in the \BM\ model. However, he assumed that the effect of the
perturbing field on the transition rates defining the trap model
dynamics has a simple multiplicative form. This is ``not easy to
justify {\em a priori}''~\cite{Ritort03} and gives only an
approximation to the natural prescription where the effect of the
perturbing field is to shift all energies according to $E\to E-hm$. I
show in this paper that the response in this natural \BM\ model can
be analysed directly, and I give exact results for the FD relations in
the limit of low temperatures; these differ in important respects from
those obtained with the approximation of multiplicatively perturbed
rates. As a by-product of the calculation, I also show briefly that
the FD relation due to Bouchaud and Dean~\cite{BouDea95}, which was
recovered by Ritort~\cite{Ritort03} for the \BM\ model with the
approximation of multiplicatively perturbed rates, is in fact valid
for {\em arbitrary} trap models with multiplicatively perturbed rates.

In the following section I give the definitions of the Bouchaud and
the \BM\ trap models; general expressions for correlation and response
functions that apply to all trap models are then derived in
Sec.~\ref{sec:corr}. Sec.~\ref{sec:mult} applies these to the case of
multiplicatively perturbed rates. Sec.~\ref{sec:BM} contains the main
result, the exact low-$T$ FD relation for the \BM\ model. I conclude
in Sec.~\ref{sec:conclusion} with a discussion of the intriguing links
between trap models with energetic and entropic barriers which arise
as a consequence of this relation.

\section{Trap models}

A trap model is defined by a distribution $\rho(E)$ of trap energies;
the convention for the sign of $E$ is here that lower $E$ corresponds
to {\em deeper} traps, which is the reverse of that in \eg~\cite{BouDea95}.
The primary dynamical quantity is then $P_0(E,t)$, the distribution of
finding the particle in a trap of energy $E$ at time $t$; the
subscript 0 indicates that for now we are considering the dynamics
without any perturbing fields. The evolution of $P_0$ is given by
the master equation
\be
\dd{t}P_0(E,t) = - \Gamma_0(E) P_0(E,t) + \rho(E) \int\! dE'\, w_0(E\ar
E')P_0(E',t)
\label{master}
\ee
where $w_0(E\ar E')$ is the rate for transitions between traps of
energy $E'$ and $E$. More precisely, if one considers a finite number
of traps $N$, the transition rate from trap $i$ to $j$ is
$(1/N)w_0(E_j\ar E_i)$; the total rate for transitions to traps in the
energy range $E<E_j<E+dE$ is then $w_0(E\ar E_i)$ times the fraction
of traps in this range, which is $\rho(E)dE$ for large $N$. The
quantity
\be
\Gamma_0(E) = \int\! dE' \rho(E') w_0(E'\ar E)
\label{Gam0}
\ee
in\eq{master} is the total ``exit rate'' from a trap of energy $E$.

Two specific instances of trap models have received considerable
attention in recent years. Bouchaud~\cite{Bouchaud92} chose for his trap
model $\rho(E) = \Tg^{-1}\exp(E/\Tg)$ with $-\infty<E<0$. For any choice of
transition rates that satisfies detailed balance, the model then has a
glass transition at $T=T_g$ since the equilibrium distribution $P_{\rm
eq}(E)\propto \rho(E)\exp(-E/T)$ becomes unnormalizable there. For
lower $T$, the system must show aging, \ie\ a strong dependence of its
properties on the waiting time $\tw$. Bouchaud~\cite{Bouchaud92}
assumed transition rates
\be
w_0(E'\ar E)=\exp(\beta E)
\label{Bouchaud}
\ee
that are independent of the energy of the
arrival trap; here $\beta=1/T$ as usual. \BM~\cite{BarMez95} chose instead
Glauber rates
\be
w_0(E'\ar E)=\frac{1}{1+\exp[\beta(E'-E)]}
\label{Glauber}
\ee
As emphasized by Ritort~\cite{Ritort03}, the out-of-equilibrium
dynamics of these models is rather different: in the Bouchaud model
with its activated dynamics, glassiness arises from the presence of
{\em energy} barriers, and the system arrests completely for $T\to
0$. In the \BM\ case, on the other hand, the system can keep evolving
by transitions to traps with ever lower energies, even at $T=0$; the
diminishing number of such traps effectively creates {\em entropic}
barriers that slow the relaxation.

\section{Correlation and response}
\label{sec:corr}

We now want to consider the correlation and response properties of
some, essentially arbitrary, observable $m$. In the most general terms
the properties of this are described by the distributions $\rho(m|E)$
of $m$ across traps of given $E$. I will assume throughout that $m$ is
on average uncorrelated with $E$, so that its conditional mean
\be
0=\int\! dm\, m\,\rho(m|E)
\label{zeromean}
\ee
vanishes for all $E$; the variance
\be
\var = \int\! dm\, m^2 \rho(m|E)
\ee
however can be dependent on $E$. With $m$ included, the master
equation is
\bea
\fl\dd{t} P(E,m,t) &=& -\Gamma(E,m)P(E,m,t)
\nn
& &{}+{} \rho(m|E) \rho(E) \int\! dE'\,dm'\, w(E,m\ar E',m')P(E',m',t)
\label{master2}
\eea
where the rates $w(E,m\ar E',m')$ may now depend on a perturbing field
$h$ conjugate to $m$, and the total exit rates are
\be
\Gamma(E,m) = \int\! dE'\,dm'\, \rho(m'|E') \rho(E') w(E',m'\ar E,m)
\label{GamEm}
\ee

An expression for the {\em correlation function} of $m$ is easily
found. In the absence of a field, $w(E',m'\ar E,m)=w_0(E'\ar E)$ and
$\Gamma(E,m)=\Gamma_0(E)$ are independent of the value of our
observable. Equation\eq{master2} then shows that
$P(E,m,t)=\rho(m|E)P_0(E,t)$ as long as the same is true at time
$t=0$. (This is a natural assumption and holds \eg\ when $P(E,m,0)$ is
an equilibrium distribution at zero field and some initial temperature
above $\Tg$, from which the system is quenched to $T<\Tg$ at $t=0$.)
For our zero mean observables\eq{zeromean} this implies in particular
that $\lav m(t)\rav=0$ at all times. The two-time correlator of $m$ is
then
\bea
\fl C(t,\tw) &=& \lav m(t)m(\tw)\rav \\
&=& \int\! dE\,dm\,dE'\,dm'\, m\,m'
P_0(E,m|E',m',t-\tw) \rho(m'|E')P_0(E',\tw)
\label{corr1}
\eea
Here $P_0(E,m|E',m',t-\tw)$ is the propagator, \ie\ the
probability of being in a trap with energy $E$ and observable $m$
when starting from a trap with $E'$ and $m'$ a time $t-\tw$
earlier. This can be obtained as the solution to\eq{master2} starting
from the initial condition $\delta(E-E')\delta(m-m')$. Since the
correlation function is calculated in the absence of a field, the only
nontrivial $m$-dependence in\eq{master2} arises from the factor
$\rho(m|E)$. Treating the second term on the r.h.s.\ of\eq{master2} as
an inhomogeneity one thus sees that
\be
\fl P_0(E,m|E',m',t-\tw) =
e^{-\Gamma_0(E')(t-\tw)}\delta(E-E')\delta(m-m') +
\rho(m|E) \times \ldots
\label{propagator}
\ee
where the dots indicate factors not involving $m$. Inserting
into\eq{corr1} and using the zero-mean assumption\eq{zeromean} then
yields the simple representation
\be
C(t,\tw) = \int\! dE\, \var e^{-\Gamma_0(E)(t-\tw)} P_0(E,\tw) 
\label{corr}
\ee
for the correlation function. This makes sense: physically, every hop
completely decorrelates the observable, so that $C$ is an average of
the probabilities $\exp[-\Gamma_0(E)(t-\tw)]$ of remaining in the
current trap, weighted by the probability of being in a trap of energy
$E$ at time $\tw$ and multiplied by the variance of $m$ across traps
of this energy.

To find the {\em impulse response} $R(t,\tw)$, consider a field impulse
of amplitude $h$
and infinitesimal length $\dt$, applied at time $\tw$. Denote
\be
\Delta w(E',m'\ar E,m) = w(E',m'\ar E,m) - w_0(E'\ar E)
\ee
the change in the transition rates caused by the field, and
$\Delta\Gamma(E,m)$ similarly the change in the total exit rates;
$h$-dependences are not written explicitly here. Then from the master
equation\eq{master2}, and using that $P(E,m,\tw)=\rho(m|E)P_0(E,\tw)$,
one has
\bea
\fl P(E,m,\tw+\dt)&=&\rho(m|E)P_0(E,\tw) \nn
& &{}-{}\dt\Gamma_0(E)\rho(m|E)P_0(E,\tw)\nn
& &{}+{}\dt \rho(m|E) \rho(E) \int\! 
dE'\,dm'\, w_0(E\ar E')\rho(m'|E')P_0(E',\tw)
\label{perturb}\\
& &{}-{}\dt\Delta\Gamma(E,m)\rho(m|E)P_0(E,\tw) \nn
& &{}+{}\dt \rho(m|E) \rho(E) \int\! 
dE'\,dm'\, \Delta w(E,m\ar E',m')\rho(m'|E')P_0(E',\tw)
\nonumber
\eea
where the effects of the field have been explicitly separated off in
the last two lines. After time $\tw+\dt$, when the field is switched
off again, the same argument that lead to\eq{propagator} applies and so
\be
P(E,m,t) = e^{-\Gamma_0(E)(t-\tw)}P(E,m,\tw+\dt) +
\rho(m|E)\times\ldots
\label{aux}
\ee
for $t>\tw+\dt$ with the dots again indicating factors independent of
$m$; in the exponent I have approximated $t-\tw-\dt\approx t-\tw$
since we are interested in the limit $\dt\to 0$. To find $\lav
m(t)\rav$, from which the response function is obtained, one
inserts\eq{perturb} into\eq{aux}, multiplies by $m$ and integrates
over $m$ and $E$. All terms of the form $\rho(m|E)\times\ldots$ give a
vanishing contribution due to\eq{zeromean}. Only the last two lines
of\eq{perturb} thus survive, and the two-time response function can be
written as
\bea
\fl hR(t,\tw) &=& \frac{1}{\dt}\lav m(t)\rav \\
&=& \int\!dE\,dm\, e^{-\Gamma_0(E)(t-\tw)}m
\biggl[-\Delta\Gamma(E,m)\rho(m|E)P_0(E,\tw) \nn
& &{}+{}\rho(m|E) \rho(E) \int\! 
dE'\,dm'\, \Delta w(E,m\ar E',m')\rho(m'|E')P_0(E',\tw)\biggr]
\label{resp1}
\eea
So far this applies for arbitrary field amplitude $h$, so that
$R(t,\tw)$ is in general a nonlinear response function, but we will
specialize to the linear response limit $h\to 0$ below.

\section{Multiplicatively perturbed rates}
\label{sec:mult}

To get concrete expressions for the response function one needs to
define how the field $h$ affects the transition rates. The natural
prescription is that all energies are shifted according to the value
of the observable, $E\to E-hm$ and $E'\to E'-hm'$. Before going on to
consider the more complicated case of the \BM\ model, I first briefly
review the situation in the Bouchaud model, where a simple relation
between correlation and response exists~\cite{BouDea95}. The
derivation will show that this relation actually applies rather
generally, being dependent only on the way the field affects the
transition rates.

For the Bouchaud model, shifting the energy $E\to E-hm$
in\eq{Bouchaud} gives the transition rate in the presence of a field
$w(E',m'\ar E,m)=\exp(\beta E-\beta hm) = \exp(-\beta hm)w_0(E'\ar
E)$. More generally, one can consider rates perturbed by the field
according to~\cite{BouDea95}
\be
w(E',m'\ar E,m) = e^{\beta h[(1-\zt)m'-\zt m]}w_0(E'\ar E)
\label{exp_perturb}
\ee
which reduces to the natural choice%
\footnote{
The multiplicative perturbation of rates~(\protect\ref{exp_perturb})
arises from the natural energy shift prescription $E\to E-hm$ only
for the activated rates~(\protect\ref{Bouchaud}). However, it has been
advocated also as an approximate treatment for \eg\ Glauber
rates~(\protect\ref{Glauber})~\cite{Ritort03}, and so is worth
considering for general $w_0(E'\ar E)$.
}
for $\zt=1$ but also maintains
detailed balance for other values of $\zt$. To linear order in $h$ one
then has
\be
\Delta w(E',m'\ar E,m) = \beta h[(1-\zt) m'- \zt m]w_0(E'\ar E)
\label{Dw}
\ee
and the corresponding change in the exit rates\eq{GamEm} is
\be
\Delta\Gamma(E,m) = 
\beta h\int\! dE'\,dm'\, [(1-\zt) m' -\zt m]\rho(m'|E') \rho(E') w_0(E'\ar E)
\ee
Using again the zero mean assumption\eq{zeromean}, the first term in
square brackets vanishes, giving with\eq{Gam0}
\be
\Delta\Gamma(E,m) = - \beta h\zt m \Gamma_0(E)
\label{DG}
\ee
One can now substitute\eq{Dw} -- with the arguments $(E,m)$ and $(E',m')$
interchanged appropriately -- and\eq{DG} into\eq{resp1}. Using $\int\!
dm'\, m'\rho(m'|E')=0$ and dividing by $h$ yields for the linear
response function
\bea
R(t,\tw) &=& \beta \int\!dE\,\var e^{-\Gamma_0(E)(t-\tw)}
\biggl[\zt \Gamma_0(E) P_0(E,\tw) \nn
& &{}+{} (1-\zt) \rho(E) \int\! dE'\, w_0(E\ar E') P_0(E',\tw)\biggr]
\label{exp_response}
\eea
From\eq{corr}, the first term in square brackets is seen to give
$-\beta \zt \partial C(t,\tw)/\partial t$. For the second one, one notes
from\eq{corr} and\eq{master} that
\bea
\frac{\partial C}{\partial \tw}
&=& \int\! dE\, \var e^{-\Gamma_0(E)(t-\tw)} \left[\Gamma_0(E)
P_0(E,\tw) + \frac{\partial P_0}{\partial \tw}(E,\tw)\right]
\\
&=& \int\! dE\, \var e^{-\Gamma_0(E)(t-\tw)}
\rho(E) \int\! dE' w_0(E\ar E')P_0(E',\tw)
\eea
which apart from prefactors is just the second term
in\eq{exp_response}. Thus, for {\em any} mean-field trap model with
the multiplicatively perturbed rates\eq{exp_perturb}, and any
zero-mean observable, one has the result given by Bouchaud and
Dean~\cite{BouDea95} for Bouchaud's trap model
\be
R(t,\tw) = -\beta \zt \frac{\partial C}{\partial t} + \beta (1-\zt)
\frac{\partial C}{\partial \tw}
\label{BD}
\ee
The above calculation shows that this relation holds entirely
independently of the precise form of the trap depth distribution%
\footnote{
The irrelevance of the form of $\rho(E)$ may well have been known to
the authors of Ref.~\protect\cite{BouDea95}, but was not stated
there. The version of~(\protect\ref{BD}) given in~\cite{BouDea95} is
nevertheless more limited than the one given here, since only
activated rates~(\protect\ref{Bouchaud}) and neutral observables $m$
were considered.
}
$\rho(E)$ or the transition rates $w_0(E'\ar E)$. In equilibrium,
where $C(t,\tw)$ is a function of $t-\tw$ only, it of course recovers
the usual FDT, $R(t,\tw) = \beta\,
\partial C(t,\tw)/\partial \tw$. Equation\eq{BD} applies in particular
to (zero-mean) {\em neutral} observables~\cite{FieSol02,CriRit02b},
where $m$ is completely decoupled from $E$ and therefore $\rho(m|E)$
is independent of $E$. It remains true also for more general
observables, however, as long as they have zero conditional
mean\eq{zeromean}.

\section{The \BM\ model}
\label{sec:BM}

\begin{figure}
\begin{center}
\epsfig{file=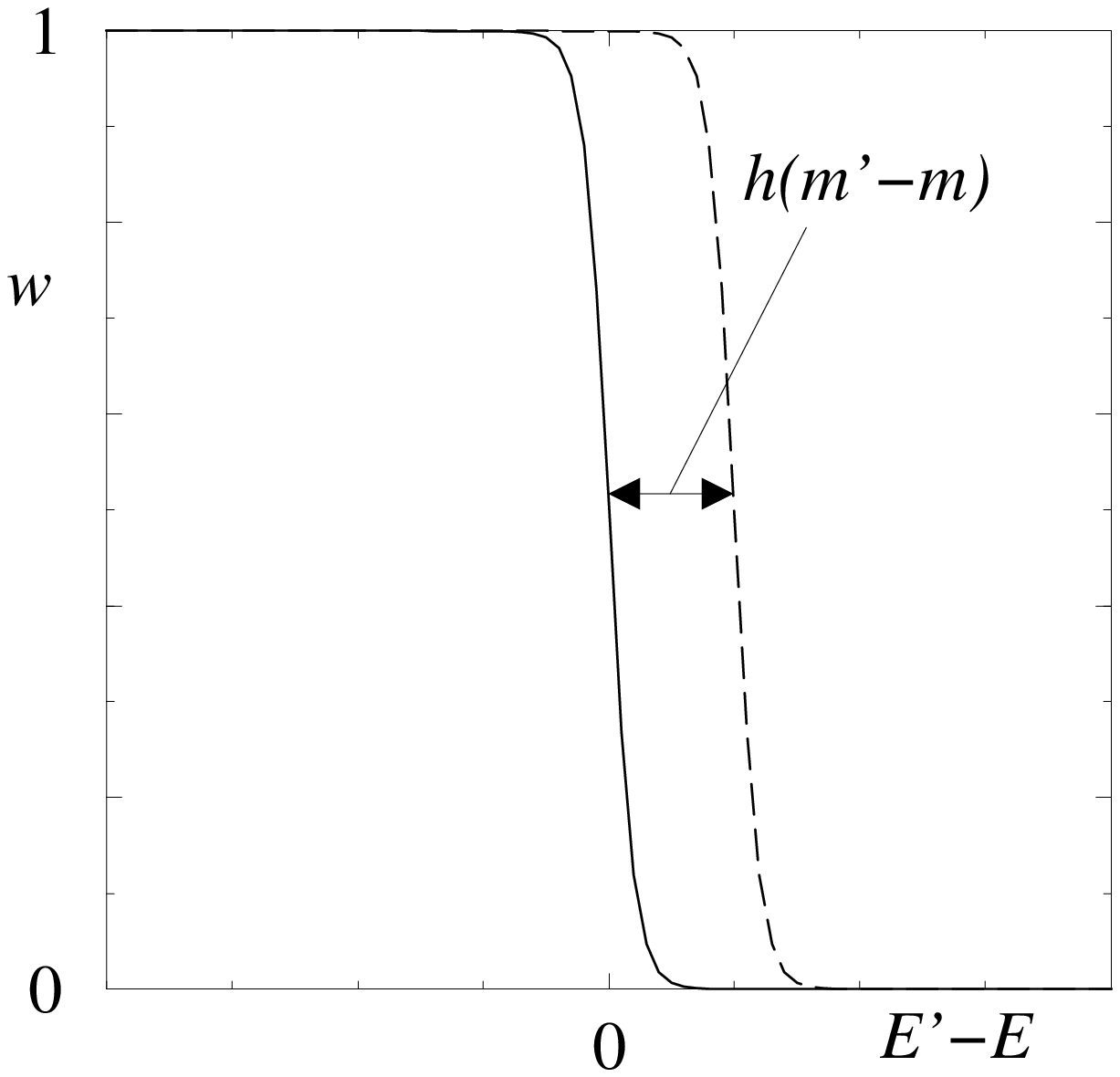,width=7cm}%
\epsfig{file=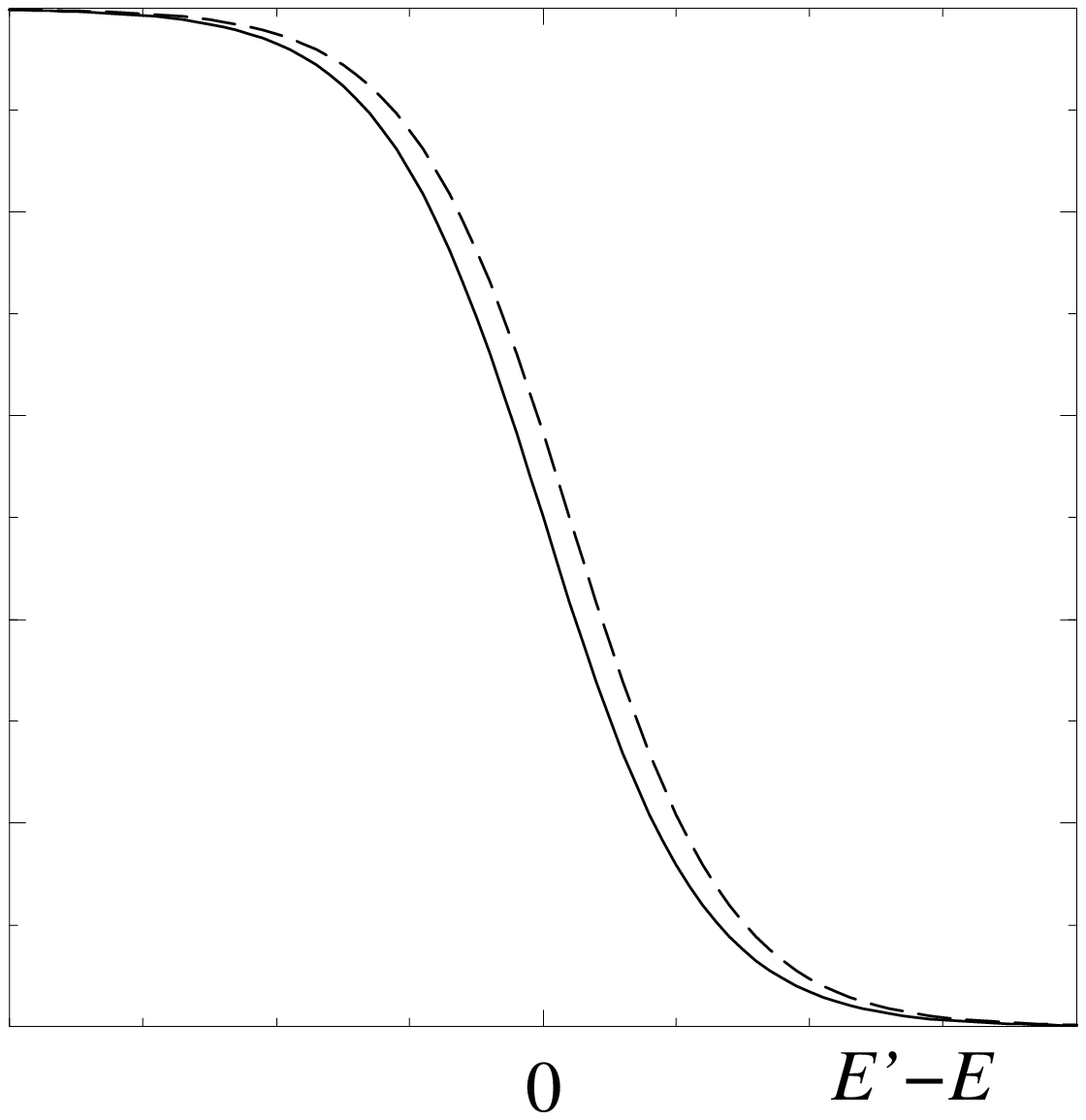,width=7cm}
\end{center}
\caption{Effect of a field on Glauber transition rates, sketched for
$h(m'-m)>0$. Solid lines show the original transition
rates~(\protect\ref{Glauber}), dashed lines those in the presence of a
field $h$, which are shifted to the right by $h(m'-m)$; see arrow on
the left. The difference between the two curves is the change in the
rates, equation\eq{rates_change}; its integral is
clearly $h(m'-m)$. Left: Case where $|h(m'-m)|>T$; the range where the
change is significant is given by $h(m'-m)$. Right: Case where
$|h(m'-m)|<T$; here the temperature $T$ sets the range where rates
change significantly. For small $h$ {\em and} small $T$ the range is small
in either case.
\label{fig1}
}
\end{figure}

Next let us turn to the \BM\ model, with the natural prescription
which assumes that the field shifts all energies. From\eq{Glauber} the
rates are then
\be
w(E',m'\ar E,m) = \frac{1}{1+\exp\{\beta[(E'-hm')-(E-hm)]\}}
\label{natural}
\ee
For low $T$, Ritort~\cite{Ritort03} argued that as a reasonable
approximation to this one could consider multiplicatively perturbed
rates
\be
w(E',m'\ar E,m) = e^{h(\gamma m'-\mu m)}w_0(E'\ar E)
\label{exp_perturb2}
\ee
with $\gamma=\mu=1/\Tg$ for exponential $\rho(E)$.
Equation\eq{exp_perturb2} is identical to\eq{exp_perturb} apart from
the replacements $\beta(1-\zt)\to \gamma$, $\beta \zt\to \mu$. As
expected from the general result\eq{BD} for multiplicatively perturbed
rates, Ritort therefore obtained the relation
\be
R(t,\tw) = -\mu\frac{\partial C}{\partial t} + \gamma
\frac{\partial C}{\partial \tw}
\label{BD_Ritort}
\ee
between response and correlation. This was found confirmed in
simulations. However, as discussed in the appendix, these simulations
were effectively performed directly with the approximate
rates\eq{exp_perturb2}, so did not give a check of how well this
approximation captures the behaviour of the \BM\ model. I now
show that the response can be calculated exactly even with the exact
rates\eq{natural}, and that the results differ from those found for
multiplicatively perturbed rates.

To calculate the response function, consider the change in the
transition rates due to the field,
\be
\fl\Delta w(E',m'\ar E,m) = \frac{1}{1+\exp\{\beta[(E'-hm')-(E-hm)]\}} -
\frac{1}{1+\exp[\beta(E'-E)]}
\label{rates_change}
\ee
This is significantly different from zero only for $E'$ within a range
of order $\max\{T,|h(m'-m)|\}$ around $E$; see figure~\ref{fig1}.  If
this range is small compared to $\Tg$, which is true for $T\ll \Tg$
and small fields $h$, then in
\be
\Delta\Gamma(E,m) =
\int\! dE'\,dm'\, \rho(m'|E') \rho(E') \Delta w(E',m'\ar E,m)
\label{change_Gamma}
\ee
we can to leading order replace $E'$ by $E$ in the factor
$\rho(E')$; the same is true for the first factor if we assume that
$\rho(m'|E')$ varies with $E'$ at most on the same scale ($\sim \Tg$)
as $\rho(E')$. Using $\int\! dE'\, \Delta w(E',m'\ar E,m) = h(m'-m)$,
which from figure~\ref{fig1} is geometrically obvious, together
with\eq{zeromean} one thus finds
\be
\Delta \Gamma(E,m) = \int\! dm'\, h(m'-m) \rho(m'|E) \rho(E) = - hm \rho(E)
\label{first}
\ee
The same argument can be applied to the integral in the second term
of\eq{resp1}, as long as we are in an out-of-equilibrium regime where
$P_0(E',\tw)$ varies with $E'$ on a scale of $\Tg$, rather than $T$ as
it would in equilibrium. This gives to leading order
\bea
\lefteqn{\int\! dE'\,dm'\, \Delta w(E,m\ar
E',m')\rho(m'|E')P_0(E',\tw) =}
\nn
&=& \int\! \,dm'\, h(m-m')\rho(m'|E)P_0(E,\tw) = h m P_0(E,\tw)
\label{second}
\eea
One can now insert\eqq{first}{second} into\eq{resp1}; after carrying
out the $m$-integration and simplifying one sees that both terms give
the same contribution. Dividing by $h$, the linear response function
is therefore
\be
R(t,\tw) = 2 \int\!dE\, \var e^{-\Gamma_0(E)(t-\tw)} \rho(E) P_0(E,\tw)
\label{resp}
\ee
Although I had implicitly assumed an exponential $\rho(E)$ above, this
result obviously remains valid also for other $\rho(E)$, as long as
$T$ is much smaller than the energy scale over which $\rho(E)$ and
$\rho(m|E)$ vary significantly. Comparing with\eq{corr}, one now sees
that in general there is no simple relation between the response and
correlation functions for the \BM\ model. However, for the exponential
trap distribution $\rho(E) = \Tg^{-1}\exp(E/\Tg)$ such a relation does
exist.  For low $T$ one can approximate the transition rates by a step
function, $w_0(E'\ar E)\approx \Theta(E-E')$ and the total exit rates
are
\be
\Gamma_0(E) = \int_{-\infty}^{E}\!\!dE'\,\rho(E') = e^{E/\Tg} = \Tg\rho(E)
\label{BMGamma}
\ee
Thus, comparing\eq{corr} and\eq{resp} gives
\be
R(t,\tw) = -\frac{2}{\Tg} \frac{\partial C}{\partial t}
\label{BM}
\ee
\begin{figure}
\begin{center}
\epsfig{file=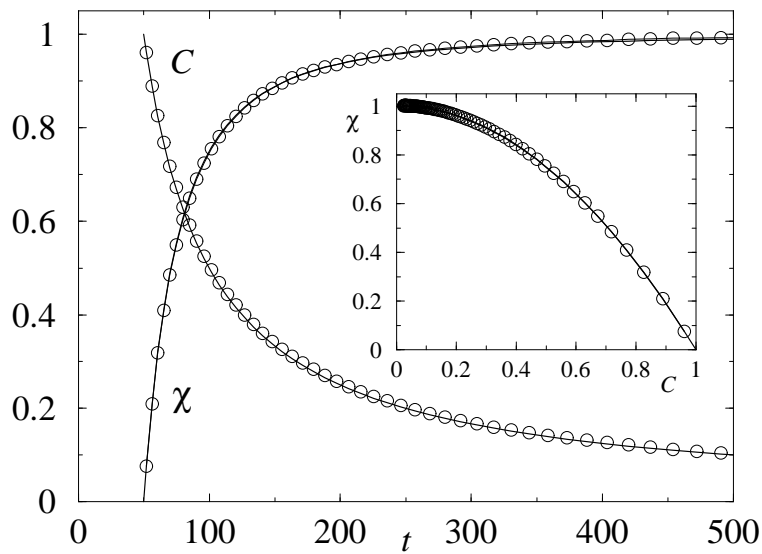,width=10cm}
\end{center}
\caption{Correlation and response of neutral observables, for the \BM\
model at $T=0$; energies are scaled such that $\Tg=1$. Main plot:
$C(t,\tw)$ and $\chi(t,\tw)$ against $t$, for $\tw=50$. The response
was determined for field $h=0.1$, which can be checked to be in the
linear regime. Circles: simulation results (see appendix). Lines:
theoretical predictions $C=\tw/t$~\protect\cite{BarMez95} and
$\chi=1-(\tw/t)^2$, see equation~(\protect\ref{parab}). Inset:
FD plot of $\chi$ vs $C$.
\label{mf_sim}
}
\end{figure}
Surprisingly, this is not dissimilar to the result\eq{BD} which one
obtains for Bouchaud's model in the most natural case $\zt=1$: the
only difference is in the prefactor, which is $1/T$ for Bouchaud's
model but $2/\Tg$ for the \BM\ model considered here.

A simple
application of\eq{BM} is to the case of a neutral observable, with
$\var=1$ (say) independently of $E$. Then from\eq{corr} one sees that
$C(t,\tw)$ is the hopping correlation function, \ie\ the probability
of not leaving the current trap between $\tw$ and $t$. This was worked
out by \BM~\cite{BarMez95} for $T\to 0$, with the result that
$C(t,\tw)=\tw/t$ for long times. Equation\eq{BM} then yields $R(t,\tw)
= (2/\Tg) \tw/t^2$; the step response follows as
\be
\chi(t,\tw) = \int_{\tw}^t\!dt'\, R(t,t')
= \frac{1}{\Tg} \left[1-\left(\frac{\tw}{t}\right)^2\right]
= \frac{1}{\Tg} \left[1-C^2(t,\tw)\right]
\label{parab}
\ee
An FD plot of $\chi$ versus
$C$ therefore has a parabolic shape, with vanishing asymptotic slope
$\partial\chi/\partial C$ for $C\to 0$, \ie\ $X_\infty=0$. The above
calculation shows that the result\eq{parab} is exact for the \BM\ model in the
limit $T\to 0$; it is also consistent with simulation results as shown
in figure~\ref{mf_sim}.

We can now assess the accuracy of the approximation of
multiplicatively perturbed rates\eq{exp_perturb2}. From\eq{BD_Ritort},
one finds in this case~\cite{Ritort03}, by arguments analogous to
those above, that $\chi = \gamma(1-C) +
\frac{\mu}{2}(1-C^2)$. Recalling that $\gamma=\mu=1/\Tg$, this is seen
to be rather different from\eq{parab}. In particular, the
approximation of multiplicatively perturbed rates incorrectly predicts
a non-vanishing asymptotic slope of the FDT plot,
$\partial\chi/\partial C=-\gamma=-1/\Tg$.

Finally, it is worth discussing a difference between the $T\to 0$
correlation and response functions for neutral observables, in terms
of their dependence of $\rho(E)$. The hopping correlation function is
{\em independent} of $\rho(E)$, as shown in~\cite{BarMez95}. Within
the formalism used here, this is clear if in\eq{corr} one sets
$\var=1$ and changes variables to the cumulative trap density
$r(E)=\int_{-\infty}^E dE'\,\rho(E')$. Together with the first part
of\eq{BMGamma} this gives $C(t,\tw)= \int_0^1\! dr\, e^{-r(t-\tw)}
P_0(r,\tw)$. Since $P_0(r,\tw)$ is independent of $\rho(E)$ (as can be
shown from\eq{master} using the same change of variable), the same
then holds for $C(t,\tw)$. The intuitive reason for this independence
is that the $T\to 0$ Glauber rates $w_0(E'\ar E)=\Theta(E-E')$ depend
only on the relative ``height'' of departure and arrival trap, but not
otherwise on the actual values of $E$ and $E'$; correspondingly, the total
exit rate $\Gamma_0(E)$ depends only on how many traps are at energies
below $E$, \ie\ on $r(E)$.

By contrast, the response function does depend on $\rho(E)$:
transforming from $E$ to $r$ in\eq{resp} gives $R(t,\tw)=2\int_0^1\!dr\,
e^{-r(t-\tw)} \rho(E(r)) P_0(r,\tw)$ and the dependence on $\rho(E)$
cannot be eliminated. This can be explained intuitively by noting that
the perturbation term $-hm$ which shifts the energies $E$ introduces
an energy scale which is not present for $h=0$. The response is
sensitive to how many traps there are with energies near (measured
on this scale) that of the departure trap, and hence to $\rho(E)$.

\section{Conclusion}
\label{sec:conclusion}

In this paper I have considered mean-field trap models, which are
simple and intuitive models of glassy dynamics. I showed briefly that
a relation between out-of-equilibrium correlation and response
functions in these models, first given by Bouchaud and Dean, is valid
for a general class of mean-field trap models; it requires only that
the transition rates are affected in the simple multiplicative
way\eq{exp_perturb} by an applied field.

I then considered the \BM\ model, which has Glauber dynamics and an
exponential distribution of trap depths. Glassiness arises in this
model from entropic barriers, rather than energetic ones as in the
Bouchaud model, and so it is of interest to compare the FD relations
that result from these different physical mechanisms.
In the natural version of the model where the
effect of a field is to shift the energies of all traps according to
the usual prescription $E\to E-hm$, the effect on the transition
rates is not simply multiplicative. The out-of-equilibrium response
can nevertheless be obtained exactly for low $T$, and one finds 
a relation which is quite similar to, but distinct from, that given by
Bouchaud and Dean. The exact calculation also shows that an
approximate treatment using multiplicatively perturbed
rates~\cite{Ritort03} gives qualitatively incorrect results.

Comparing the above results for the (natural) \BM\ model with those
for Bouchaud's model (with, likewise, the natural choice $\zt=1$), one
notes two intriguing parallels for the low-temperature
out-of-equilibrium dynamics.  Firstly, both models give FD plots with
$X_\infty=0$, \ie\ with a slope $\partial \chi/\partial C$ which tends
to {\em zero} in the limit $C\to 0$. (For the \BM\ model with non-neutral
observables%
\footnote{
Strictly speaking, as shown in~\protect\cite{Ritort03}, one requires
observables that probe only the aging dynamics, in the sense that
their correlation function only decays on timescales that grow with
$\tw$. A counterexample would be an observable that is sensitive only
to the very shallow traps, which in Bouchaud's model gives a
correlation function that decays completely on timescales of
$\order(1)$~\protect\cite{FieSol02}.
}
this can be deduced by applying the arguments of~\cite{Ritort03} to the
relation\eq{BM}.)  Second, the value of the susceptibility itself in
the same limit is $\chi_\infty=1/\Tg$ in both models for neutral
observables; see again\eq{parab}. This is precisely the value that one
would expect if, as $T$ is lowered, $\chi_\infty$ ``freezes'' at
$T=\Tg$ and remains independent of $T$ for $T<\Tg$. For Bouchaud's
model this $T$-independence can indeed be shown~\cite{Ritort03}; for
the \BM\ model the result $\chi_\infty=1/\Tg$ found above for $T\to 0$
strongly suggests that $\chi_\infty$ is likewise $T$-independent for
$0<T<\Tg$. Even though the slow out-of-equilibrium dynamics in the two
models is very different, being caused by activation over energy
barriers for Bouchaud's model and by entropic barriers for the \BM\
model, we thus have the intriguing observation that
some features of the out-of-equilibrium FD relations are shared. It
will be interesting to explore whether this correspondence extends to
other properties, and possibly to other models of glassy dynamics.

\ack I warmly thank Felix Ritort for rekindling my interest in the
\BM\ model, for informing me of his work~\cite{Ritort03} before
submission, and for enlightening discussions. Financial support from
the Nuffield Foundation through award NAL/00361/G is gratefully
acknowledged.

\section*{Appendix: Simulation method}

To simulate any mean-field trap model, one can use that in the limit
$N\to \infty$ no trap is visited twice, so that $E$ and $m$ can be
sampled anew at each transition and no explicit population of traps
needs to be maintained. The probability for making a transition from a
trap with $(E,m)$ to one with $(E',m')$ is
\be
P_1(E',m'|E,m) = \Gamma^{-1}(E,m) \rho(m'|E')\rho(E') w(E',m'\ar E,m)
\label{aux1}
\ee
and contains $\Gamma(E,m)$, the total exit rate from the current trap,
as a normalization factor; see\eq{GamEm}. Now specialize to the \BM\
model, with $\rho(E)=\Tg^{-1}\exp(E/\Tg)$, $E<0$ and a neutral observable for
which I take $\rho(m|E)\equiv \rho(m)$ as a zero mean, unit variance
Gaussian independently of $E$. The transition rates at $T=0$ are
$w(E',m'\ar E,m)=\Theta(E-hm-E'+hm')$. Integrating over $E'$
in\eq{aux1} then gives
\be
P_1(m'|E,m) = \Gamma^{-1}(E,m)\rho(m') e^{(E-hm+hm')/\Tg}
\propto \rho(m')e^{hm'}
\label{aux2}
\ee
Dividing\eq{aux1} by this yields
\be
P_1(E'|m',E,m) = \Theta(E-hm+hm'-E') \Tg^{-1} e^{(E'-E+hm-hm')/\Tg}
\label{aux3}
\ee
which is just an exponential distribution over
$-\infty<E'<E-hm+hm'$. One can thus sample from\eq{aux1} by first
sampling $m'$ from\eq{aux2}, which is a Gaussian with unit variance
and mean $h$; after that one samples $E'$ from\eq{aux3}. The total
exit rate follows \eg\ from normalization of\eq{aux2} as
\be
\Gamma(E,m)=\exp[(E-hm)/\Tg+h^2/(2\Tg^2)]
\label{aux4}
\ee

It is important to note from\eq{aux3} that the distribution of $E'$
depends on $m-m'$. One might be tempted to neglect this dependence,
replacing\eq{aux3} by $\Theta(E-E')\Tg^{-1}
e^{(E'-E)/\Tg}$~\cite{Ritort03}. However, by repeating the
calculations leading to\eqqq{aux2}{aux3}{aux4} one easily checks that
this is equivalent to changing from the exact rates\eq{natural} to the
multiplicatively perturbed rates\eq{exp_perturb2} with
$\gamma=\mu=1/\Tg$. As shown above, this leads to rather
different response functions; the precise form of\eq{aux3} is thus
important to get the correct results.

The results shown in figure~\ref{mf_sim} were obtained for a quench
from $T=\infty$ at $t=0$, corresponding to the initial condition
$P(E,m,0)=\rho(m)\rho(E)$, and averaged over $5\times 10^7$
runs. Direct simulations with a population of $N=10^8$ traps yielded
equivalent results, though one needs to be aware of finite-$N$ effects
which become more acute for low $E$ because of the exponential
decrease in the population density $\rho(E)$.

\section*{References}


\end{document}